\documentclass[nopreprintline,review,12pt]{elsarticle}

\usepackage{graphicx}
\usepackage{subcaption}
\usepackage{array}
\usepackage{tabularx}
\usepackage{amsmath}
\usepackage{setspace}
\usepackage{multirow}

\DeclareMathOperator*{\argmax}{argmax}

\newcommand{\PreserveBackslash}[1]{\let\temp=\\#1\let\\=\temp}
\newcolumntype{C}[1]{>{\PreserveBackslash\centering}p{#1}}

\begin{document}

\begin{frontmatter}



\title{\vspace*{-2.5cm}Motion Magnification Algorithms for Video-Based Breathing Monitoring}


\author[add1]{Veronica Mattioli\corref{cor1}}
\cortext[cor1]{Corresponding author}
\ead{veronica.mattioli@unipr.it}
\author{Davide Alinovi\fnref{label2}}
\fntext[label2]{Deceased on 16 September 2020}
\author[add1]{Gianluigi Ferrari}
\ead{gianluigi.ferrari@unipr.it}
\author[add2]{Francesco Pisani}
\ead{francesco.pisani@uniroma1.it}
\author[add1]{Riccardo Raheli}
\ead{riccardo.raheli@unipr.it}

\address[add1]{Department of Engineering and Architecture, University of Parma, Parco Area delle Scienze 181/A, 43124 Parma, Italy}
\address[add2]{Department of Human Neuroscience, Sapienza University of Rome, Viale dell'Università 30, 108, 00185 Rome, Italy}


\begin{singlespace}
\begin{abstract}
In this paper, we present two video processing techniques for contact-less estimation of the Respiratory Rate (RR) of framed subjects. Due to the modest extent of movements related to respiration in both infants and adults, specific algorithms to efficiently detect breathing are needed. For this reason, motion-related variations in video signals are exploited to identify respiration of the monitored patient and simultaneously estimate the RR over time. Our estimation methods rely on two motion magnification algorithms that are exploited to enhance the subtle respiration-related movements. In particular, amplitude- and phase-based algorithms for motion magnification are considered to extract reliable motion signals. The proposed estimation systems perform both spatial decomposition of the video frames combined with proper temporal filtering to extract breathing information. After periodic (or quasi-periodic) respiratory signals are extracted and jointly analysed, we apply the Maximum Likelihood (ML) criterion to estimate the fundamental frequency, corresponding to the RR. The performance of the presented methods is first assessed by comparison with reference data. Videos framing different subjects, i.e., newborns and adults, are tested. Finally, the RR estimation accuracy of both methods is measured in terms of normalized Root Mean Squared Error (RMSE).
\end{abstract}

\begin{keyword}
respiratory rate estimation, motion magnification, maximum likelihood, video processing
	
	
	
\end{keyword}

\end{singlespace}
%
%

\end{frontmatter}


\section{Introduction}\label{sec_intro}
Breathing monitoring is a fundamental diagnostic tool to assess the physiological status of a patient. In particular, the Respiratory Rate (RR) is a relevant indicator of potential human respiratory system dysfunctions that may be caused by critical medical conditions. Typical values of the RR in healthy adults at rest lie between 12 and 20 breaths per minute and may vary with age. The RR in newborns and children is usually higher. Abnormal values of the RR may be a sign of severe issues arising from respiratory disorders or complications. For instance, diseases such as chronic obstructive pulmonary disease, asthma, anaemia and epileptic seizures may cause oxygen levels in the blood to significantly drop, potentially leading to cyanosis, cerebral palsy or cardiac arrest and ischaemic events \cite{costanzo_review}. 

A constant and careful monitoring of the respiration of a patient is crucial for early diagnosis and intervention that may be lifesaver in some cases. Recent and extensive reviews about current RR monitoring methodologies can be found in~\cite{costanzo_review, massaroni_review}. In particular, these monitoring systems are typically classified into two main categories: contact-based and contact-less. In~\cite{costanzo_review}, a thorough analysis of both categories is presented, whereas~\cite{massaroni_review} focuses on contact-less methods only, that are gaining increasing attention thanks to the advantages they may provide. In fact, they are particularly suitable for remote monitoring, that has become fundamental especially in the pandemic era in which patients affected by COVID-19 need constant medical attention~\cite{massaroni_covid}. Furthermore, contact-less devices include Red, Green and Blue (RGB) and Infra Red (IR) cameras, as well as microphones, among other sensors \cite{massaroni_review}, whose cost is significantly lower than sophisticated equipments, usually deployed in hospital environments. These instruments are also non-invasive, hence more comfortable, as they do not require a direct contact with the body of the patient. On the other hand, contact-based methods include more invasive procedures, such as the pneumography~\cite{FREUNDLICH1974181} and phlebotomy~\cite{phlebo}. The former technique allows to measure the thoracic movements by means of an elastic belt placed around the chest of the patient, whereas the latter allows to sample arterial, capillary or venous blood gas. Despite its high accuracy, phlebotomy may be painful and difficult to perform, especially in children and newborns, and may lead to complications such as thrombosis, haemorrhage and aneurysm formation~\cite{phlebo}.

Other conventional probes to monitor the cardiac and pulmonary activity are the ElectroCardioGram (ECG) and the Pneumogram that require wired electrodes to be directly attached to the chest of the patient. The main limitation of these instruments is their deployment, as it is mainly limited to clinical settings, being not suitable for home care. As another example of contact-based devices, we mention the pulse oximeter, that has become very popular nowadays as it allows to easily measure the oxygen saturation in the blood, also in domestic environments. For example, this is, indeed, a very informative parameter about the severity of the COVID-19 disease. The pulse oximeter is usually clipped to the fingertip of a patient and measures the changes in the transmission or reflection of the light emitted by a  Light-Emitting Diode (LED) hitting the skin of the subject. This working principle is referred to as PhotoPlethysmoGraphy (PPG) and has also inspired some contact-less video-based monitoring methodologies, as discussed in the following. \\ 

\vspace*{-0.4cm}Among contact-less methodologies for the RR monitoring, video processing systems are becoming very appealing. Three main approaches can be identified based on (a)~PPG, (b)~optical flow and (c)~motion magnification~\cite{massaroni_review}.

In \cite{massaroni_ppg}, the respiratory signal is extracted from selected PPG signals computed on a Region Of Interest (ROI) that surrounds the pit of the neck of the subject. The RR is estimated in frequency and time domains and the performance of different RGB camera sensors is analysed. The PPG principle is also exploited in \cite{userface}, but the hue channel of the Hue, Saturation and Value (HSV) colour space is considered for the analysis. On the other hand, the optical flow may be exploited to detect and track breathing-related movements, as in~\cite{MATEUMATEUS2021102443, janssen}. However, since respiration movements may be subtle and difficult to detect, especially in newborns, motion magnification techniques may be applied to enhance them, as in~\cite{cattani, ALNAJI20161, stabilized_evm}.

An effective mathematical model of the RR and its possible disorders is presented in \cite{ALINOVI2017245}. It is based on a time-continuous Markov chain and enables the implementation of video-based simulations of breathing disorders which may be useful in the design of RR estimation algorithms.


This paper analyses motion magnification algorithms for RR estimation. Amplitude- and phase-based techniques, respectively inspired by the works in~\cite{evm1} and~\cite{rubinstein_riesz}, are considered. In particular, in \cite{evm1} spatial and temporal processing is combined to amplify the variations of the pixel intensities for frequency bands of interest. In~\cite{rubinstein_riesz}, an approximation of the Riesz transform is proposed to perform phase amplification of motion signals. Unlike the approaches in these references, video reconstruction with amplified motion is not necessary in breathing monitoring. Once the amplified motion signal components are obtained, an estimation technique, based on the Maximum Likelihood (ML) principle~\cite{Kay}, can be applied to estimate the RR. This paper expands upon preliminary conference contributions~\cite{memea2015, isspit2016, eusipco2018}.

The related work in~\cite{ALNAJI20161} is based on the method proposed in~\cite{evm1}, but performs amplitude magnification and the unnecessary final video reconstruction, as also done in~\cite{cattani}. The work in~\cite{stabilized_evm}, based on the phase magnification technique described in~\cite{rubinstein_riesz}, presents a method for video stabilization for handheld cameras and the procedure to extract useful phase signals in this context.  

The remainder of the paper is organized as follows. In Section \ref{sec_motionsignal}, the extraction of the motion signals is described and two methods for motion magnification are detailed. In Section \ref{sec_ml}, the RR estimation procedure is presented and the operation of automatically selecting ROIs where breathing-related movements are present is described along with a decision strategy to discard unsuitable ROIs. In Section \ref{sec_res}, the performance of the considered methods is discussed and compared against reference data. Finally, in Section \ref{conclusion} conclusions are drawn. 

\section{Motion Signal Extraction}\label{sec_motionsignal}
In this section, the extraction of motion signals from video sequences is described. Full-frame video sequences are initially considered as inputs to the proposed methods for an effective and simple description. Nevertheless, ROIs can also be extracted to reduce the computational complexity as well as to improve robustness. To this end, a method to automatically select ROIs will be presented in Subsection~\ref{sec_roi}.

In the following, we refer to a generic gray scale video sequence, acquired with a sampling rate $f_{s}$ (dimension: [frame/s]), as a discrete signal $f[\mathbf{u},n]$ that defines the pixel intensities at position $ \mathbf{u}=(u_{1}, u_{2}) $ at the $n$-th frame. Each frame has size $U_{1} \times U_{2}$ (dimension: [pixel]) and is sampled at time instants $nT_{s}$ (dimension: [s]), where $T_{s} = 1/f_{s}$ is the sampling period. The videos considered in this paper, recorded by RGB cameras, can be converted to gray scale~\cite{Solomon}.

\subsection{Amplitude-based Motion Magnification}\label{sec_ampl}
\begin{figure}[t!]                                                                                                                                                   
	\hspace*{-0.4cm}
	\centering
	\includegraphics[width=1.05\textwidth]{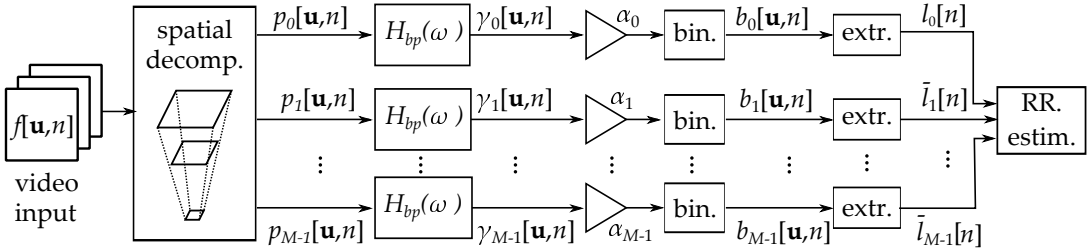}
	\caption{Amplitude-based (spatio-temporal) RR estimation algorithm.}
	\label{stvp}
\end{figure}

Amplitude-based techniques for motion magnification aim at linearly amplifying variations of each pixel intensity over time. The method proposed in~\cite{evm1}, called Eulerian Video Magnification (EVM), performs temporal processing on different spatial frequency bands obtained by decomposing each frame of the input video into a set of subimages. The processed and unprocessed video subsignals so obtained are finally recombined to obtain the amplified output video. Unlike the preliminary work in~\cite{cattani}, we present here a spatio-temporal approach to extract useful motion signals, inspired by the EVM algorithm in~\cite{evm1}, in which this final recombination step is not performed because not of interest for the purpose of breathing monitoring. An illustrative overview of the proposed method is shown in Figure~\ref{stvp}, where each processing step is associated with a diagram block and is detailed hereafter.

\paragraph{Spatial decomposition}
As a first step, each frame of the video $f[\mathbf{u},n]$ is decomposed into a set of $M$ subimages with scaled resolutions, each representing a different spatial frequency band. The $M$ scaled subimages, referred to as ``levels'', are obtained by computing a Laplacian pyramid \cite{laplacian} and are sorted with decreasing resolution. A Laplacian pyramid is formed as follows. First, a Gaussian pyramid \cite{laplacian} is derived, where $g_{0}[\mathbf{u},n] = f[\mathbf{u},n] $ is set as the bottom level, that corresponds to the highest spatial frequency band and is characterized by the highest resolution. Upper levels, representing lower spatial frequency bands and characterized by lower resolutions, are recursively computed according to a ``reduce'' function defined as
\begin{equation}
	g_{m}[\mathbf{u},n] = \sum_{k_{1} = - R_{M}}^{+R_{M}} \sum_{k_{2}= - R_{M}}^{+R_{M}} w[k_{1},k_{2}] g_{m-1}[2u_{1}-k_{1}, 2u_{2}-k_{2} ,n]
	\label{eq_reduce}
\end{equation}
where $m = 1, \dots M-1$ denotes the $m$-th pyramid level, $ w[k_{1},k_{2}] $ is a proper truncated Gaussian low-pass filter, designed according to specific constraints described in \cite{laplacian}, and $ R_{M} $ is a positive integer that specifies the size of this filter as $(2R_{M} + 1) \times (2R_{M} + 1)$. An ``expand'' function can also be defined as
\begin{equation}
	\hat{g}_{m}[\mathbf{u},n] = 4\sum_{k_{1}= - R_{M}}^{+R_{M}} \sum_{k_{2}= - R_{M}}^{+R_{M}} w[k_{1},k_{2}] g_{m+1}\bigg[\dfrac{u_{1}-k_{1}}{2}, \dfrac{u_{2}-k_{2}}{2} ,n\bigg]
	\label{eq_expand}
\end{equation}
to obtain a specific level by expanding the dimensions of the upper one (with lower resolution) by interpolation. The filter mask $ w[k_{1},k_{2}] $ is the same in (\ref{eq_reduce}) and (\ref{eq_expand}).

The Laplacian pyramid levels are derived from (\ref{eq_reduce}) and (\ref{eq_expand}) as
\begin{equation}
p_{m}[\mathbf{u},n] =
\begin{cases}
g_{m}[\mathbf{u},n] - \hat{g}_{m}[\mathbf{u},n] & m = 1, \dots, M-2\\
g_{m}[\mathbf{u},n] & m = M-1
\end{cases}
\label{eq_laplacian}
\end{equation}
where $p_{M-1}[\mathbf{u},n] = g_{M-1}[\mathbf{u},n]$ is set as the highest-index level and describes the lowest spatial frequency band. The expression in (\ref{eq_laplacian}) represents the error image between a level of the Gaussian pyramid $g_{m}$ and the same level $\hat{g}_{m}$ obtained by expanding the upper one according to the function in (\ref{eq_expand}).

The operation of spatial decomposition is highlighted in the first block of the diagram in Figure \ref{stvp}.

\begin{figure}[t!]
	\centering
	\includegraphics[width=0.9\textwidth,trim=1cm 5cm 3cm 0cm]{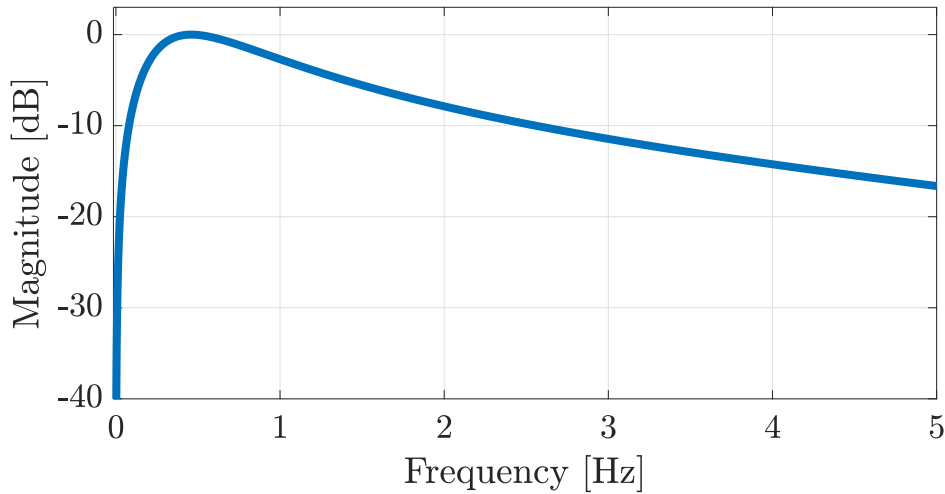}
	\caption{Frequency response of the IIR filter employed for adults.}
	\label{filter}
\end{figure}

\paragraph{Temporal filtering}
Once the spatial processing is performed and a spatial decomposition based on the Laplacian pyramid is obtained, each level is pixel-wise temporally filtered to extract a frequency band that corresponds to a typical range of RR. A Butterworth filter of the second order with Infinite Impulse Response (IIR) can be selected as a proper temporal digital band-pass filter. Its transfer function can be expressed as
\begin{equation}
	H_{bp}(z) = K \dfrac{(1+z^{-1})(1-z^{-1})}{(1-pz^{-1})(1-p^{*}z^{-1})}
	\label{eq_butter}
\end{equation}
where the scale factor $K$ and the complex conjugates poles $p$ and $p^{*}$ can be computed following the filter design rules to fit the requirements for the lower and upper 3-dB cut-off frequencies $f_{\mathrm{L}}^{\mathrm{co}} $ and $f_{\mathrm{H}}^{\mathrm{co}}$~\cite{Oppenheim}. In this work, the cut-off frequencies of the filter are set according to to the framed subject: for adults $f_{\mathrm{L}}^{\mathrm{co}} = 0.19$~Hz and $f_{\mathrm{H}}^{\mathrm{co}} = 0.9$~Hz, corresponding to a range of $11-54$~breath/min, whereas for newborns $f_{\mathrm{L}}^{\mathrm{co}} = 0.3$ Hz and $f_{\mathrm{H}}^{\mathrm{co}} = 1.1$~Hz, corresponding to a range of $18-66$~breath/min. The frequency response of the IIR filter employed for adults is shown in Figure~\ref{filter}.

The temporal processing is represented as a filter bank in Figure~\ref{stvp} and the obtained filtered levels are denoted as $\{\gamma_{m}[\mathbf{u},n]\}_{m = 0}^{M-1}$.

\paragraph{Signal amplification}
Each filtered level $\gamma_{m}[\mathbf{u},n]$, $m = 0, \dots, M-1$, is multiplied by a proper amplification factor to linearly amplify motions related to the respiration. The amplification coefficients are denoted as $ \{\alpha_{m}\}_{m = 0}^{M-1} $ in Figure~\ref{stvp} and are properly set according to \cite{evm1} to avoid noise amplification or motion artefacts. The amplification coefficient for the lowest-index level is set as $ \alpha_{0} = 1 $ and increasing values of amplification are used for higher-index levels, up to $ \alpha_{M-2} = 12 $. As the highest-index level has too low resolution to provide useful information, $ \alpha_{M-1}$ is set to 0.


\paragraph{Binarization}
Binarization is performed pixel-wise on the amplified signals $ \{\gamma_{m}[\mathbf{u},n]\alpha_{m}\}_{m = 0}^{M-1} $ to reduce the computational complexity (blocks labelled ``bin.'' in Figure~\ref{stvp}). This operation allows to highlight the respiration movements by setting to 1 the pixel intensity values larger than a preset threshold, whereas the rest of the framed scene is set to 0. This operation yields the following binarized levels, also highlighted in Figure~\ref{stvp}
\begin{equation}
	b_{m}[\mathbf{u},n] =
	\begin{cases}
	1 & \mathrm{if} \ |\{\gamma_{m}[\mathbf{u},n]\alpha_{m}\}| \geq \varGamma_{\mathrm{th}}\\
	0 & \mathrm{otherwise} 
	\end{cases}
	\label{eq_binarization}
\end{equation}
where $\varGamma_{\mathrm{th}}  $ is a proper binarization threshold heuristically set to adjust the sensitivity to motion.\footnote{The set of coefficients $\{\alpha_{m}\}_{m = 0}^{M-1}$ and the threshold $ \varGamma_{\mathrm{th}} $ in (\ref{eq_binarization}) can be scaled by a common factor without affecting the binarized frames. This lability is overcome by setting $ \alpha_{0} = 1 $.}

\paragraph{Signal extraction}
As a last step, the motion signals are extracted by spatial averaging each binarized level of the pyramid as
\begin{equation}
	\bar{l}_{m}[n] = \frac{1}{c_{m}r_{m}}\sum_{u_{1}=1}^{c_{m}} \sum_{u_{2}=1}^{r_{m}} b_{m}[\mathbf{u},n]
	\label{eq_signal}
\end{equation}
where $ \{c_{m}\}_{m = 0}^{M-1} $ and $\{r_{m}\}_{m = 0}^{M-1} $ are the widths and heights of the binarized frames (blocks labelled ``extr.'' in Figure \ref{stvp}).

\subsection{Phase-based Motion Magnification}\label{sec_phase}
Amplitude-based motion magnification presents some limitations directly linked to the linear amplification operation. When the analysed motion is small, a pixel intensity variation can be approximated by a first-order Taylor series expansion as described in~\cite{evm}. If the small motion condition is not verified, or the amplification factor $\alpha_{m}$ is too large, the approximation is not accurate and the magnification may cause undesired artefacts. Furthermore, for  $\alpha_{m} > 1$, noise is also amplified.

\begin{figure}[t!]
	\hspace*{-1.4cm}
	\centering
	\includegraphics[width=1.2\textwidth]{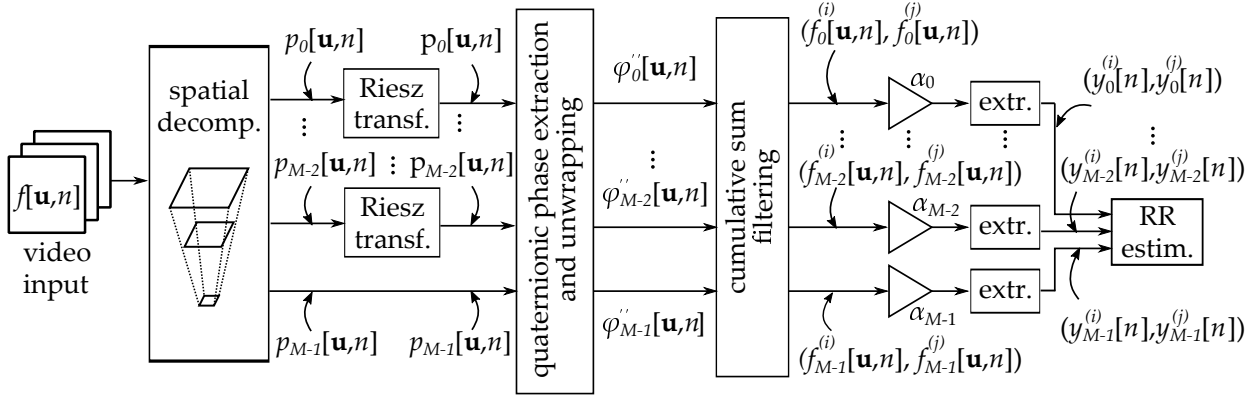}
	\caption{Phase-based RR estimation algorithm.}
	\label{phase}
\end{figure}

A solution to overcome problems related to linear amplification is provided by phase-based magnification methods, that aim at amplifying the phase of each pyramidal subsignal~\cite{rubinstein_riesz}. We present here an algorithm for motion magnification inspired by~\cite{rubinstein_riesz}, whose illustrative overview is shown in Figure~\ref{phase}, in which each processing step is associated with a diagram block and will be detailed hereafter.

\paragraph{Spatial decomposition}
Similarly to the amplitude-based (spatio-temporal) method presented in Subsection~\ref{sec_ampl}, the first step to extract amplified motion signals, as also highlighted in the first block of Figure~\ref{phase}, consists in decomposing each frame of the input video sequence $f[\mathbf{u},n]$ into a set of $M$ scaled levels by computing the Laplacian pyramid~\cite{laplacian} according to~(\ref{eq_reduce})-(\ref{eq_laplacian}). An efficient representation of the signals, where amplitudes and phases are highlighted, can now be adopted by computing the Riesz transform~\cite{monogenicsignal} of all the pyramid levels $ \{p_{m}[\mathbf{u},n]\}_{m = 0}^{M-1} $. The Riesz transform can be defined as a two-Dimensional (2D) generalization of the Hilbert transform and its 2D frequency response in the Fourier domain can be expressed as~\cite{unser}  

\begin{equation}
	H(\boldsymbol{\omega}) =
	\left( \begin{array}{c}
	H_{1}(\omega_{1}) \\ H_{2}(\omega_{2})
	\end{array} \right) =
	\left( \begin{array}{c}
	-j \omega_{1} / ||\boldsymbol{\omega}|| \\ -j \omega_{2} / ||\boldsymbol{\omega}||
	\end{array} \right)
\end{equation}
where $ \boldsymbol{\omega} = (\omega_{1},\omega_{2}) $ is the 2D vector of normalized angular frequencies and $ ||\cdot|| $ is the euclidean norm operator. The following operation, shown in the second bank of blocks in Figure \ref{phase}, is hence performed:
\begin{equation}
	\mathcal{R}\{p_{m}[\mathbf{u},n]\} = 
	\left( \begin{array}{c}
	 r_{1,m}[\mathbf{u},n] \\ r_{2,m}[\mathbf{u},n] 
	\end{array} \right) =
	\left( \begin{array}{c}
	h_{1}[\mathbf{u}]*p_{m}[\mathbf{u},n] \\ h_{2}[\mathbf{u}]*p_{m}[\mathbf{u},n]
	\end{array} \right)
	\label{eq_riesz}
\end{equation}
where $ \mathcal{R}\{ \cdot \} $ represents the Riesz transform operator, $ h_{i}[\mathbf{u}] = \mathcal{F}^{-1}(H_{i}(\boldsymbol{\omega}))$, $ i = 1,2$, $*$ denotes the 2D convolution operator and $ \mathcal{F}^{-1}(\cdot) $ is the inverse 2D Fourier transform operator.

The following triple of elements
\begin{equation}
	\mathrm{p}_{m}[\mathbf{u},n] = (p_{m}[\mathbf{u},n], r_{1,m}[\mathbf{u},n], r_{2,m}[\mathbf{u},n] )
	\label{eq_monogenic}
\end{equation}
is known as the monogenic signal of the $ m $-th level. In particular, the combination of $ \{\mathrm{p}_{m}[\mathbf{u},n]\}_{m=0}^{M-2} $ with the last level of the Laplacian pyramid $ p_{M-1}[\mathbf{u},n] $ forms a Riesz pyramid.

It may be convenient to represent the monogenic signal in (\ref{eq_monogenic}) as the quaternion \cite{rubinstein_quatern}
\begin{equation}
	\mathrm{q}_{m}[\mathbf{u},n] = p_{m}[\mathbf{u},n] + i r_{1,m}[\mathbf{u},n] + j r_{2,m}[\mathbf{u},n] + k \cdot 0
	\label{eq_quaternion}
\end{equation}
where $ i $,  $ j $ and  $ k $ are the imaginary units. Following the quaternionic algebra in \cite{rubinstein_quatern}, the norm and the natural logarithm of the quaternion in (\ref{eq_quaternion}) can be, respectively, defined as
\begin{align}
	\|\mathrm{q}_{m}\| &= \sqrt{p_{m}[\mathbf{u},n]^{2} + r_{1,m}[\mathbf{u},n]^{2} + r_{2,m}[\mathbf{u},n]^{2}} \\
	\log(\mathrm{q}_{m}) &= \log(\|\mathrm{q}_{m}\|) + \frac{i r_{1,m}[\mathbf{u},n] + j r_{2,m}[\mathbf{u},n]}{\|i r_{1,m}[\mathbf{u},n] + j r_{2,m}[\mathbf{u},n]\|}\arccos\frac{p_{m}[\mathbf{u},n]}{\|p_{m}[\mathbf{u},n]\|}
	\label{eq_qnorm}.
\end{align}
The amplitude and the quaternionic phase of (\ref{eq_quaternion}) can now be computed as
\begin{align}
	A_{m}[\mathbf{u},n] &= \| \mathrm{q}_{m}[\mathbf{u},n] \| \label{eq_ampl}  \\
	i \varphi_{m}[\mathbf{u},n] \cos(\vartheta_{m}[\mathbf{u},n]) + j \varphi_{m}[\mathbf{u},n] \sin(\vartheta_{m}[\mathbf{u},n]) &= \mathrm{log}(\mathrm{q}_{m}[\mathbf{u},n]/ \| \mathrm{q}_{m}[\mathbf{u},n] \|)
	\label{eq_phase}
\end{align}
where
\begin{align}
	\varphi_{m}[\mathbf{u},n] &= \arctan \left( \left( \sqrt{r_{1,m}[\mathbf{u},n]^{2} + r_{2,m}[\mathbf{u},n]^{2}} \right)/ p_{m}[\mathbf{u},n]  \right) \label{eq_localp}\\
	\vartheta_{m}[\mathbf{u},n] &= \arctan\left( r_{2,m}[\mathbf{u},n]/r_{1,m}[\mathbf{u},n]\right) \label{eq_localo}
\end{align}
are the $m$-th phase and orientation, respectively. The main advantage of this signal representation is that the quaternionic phase in (\ref{eq_phase}) is invariant to the signs of the phase and orientation in (\ref{eq_localp}) and (\ref{eq_localo}) \cite{rubinstein_quatern}.

\paragraph{Temporal filtering}
As a second step of the proposed phase amplification method, temporal filtering is again necessary to select a range of frequencies of interest. An IIR band-pass second-order Butterworth filter with lower and higher cut-off frequencies $f_{\mathrm{L}}^{\mathrm{co}}$ and $f_{\mathrm{H}}^{\mathrm{co}}$ can be employed to filter the phases of each level of the Riesz pyramid. As discussed in~\cite{rubinstein_quatern}, the quaternionic phases in~(\ref{eq_phase}) are first unwrapped and their cumulative sum is subsequently filtered. To this purpose, the quaternionic logarithm of the $m$-th ($m = 0, \dots, M-1$) normalized Riesz pyramid coefficient is computed~as 
\begin{equation}
\begin{cases}
	\mathrm{log}(\mathrm{\bar{q}}_{m}[\mathbf{u},n]) \ & \mathrm{for} \ n = 0 \\
	\mathrm{log}(\mathrm{\bar{q}}_{m}[\mathbf{u},n]\mathrm{\bar{q}}_{m}^{-1}[\mathbf{u},n-1]) \ & \mathrm{for} \ n = 1,2,\ldots
\end{cases}
\label{eq_coeffnorm}
\end{equation}
where $\mathrm{\bar{q}}_{m}[\mathbf{u},n] = \frac{\mathrm{q}_{m}[\mathbf{u},n]}{\|\mathrm{q}_{m}[\mathbf{u},n]\|}$ is the normalized quaternion~\cite{rubinstein_quatern} and we recall the following definitions of the inverse and conjugate quaternion, considering~(\ref{eq_quaternion})
\begin{align}
	\mathrm{q}_{m}^{-1} &= \dfrac{\mathrm{q}_{m}^{\ast}[\mathbf{u},n]}{\|\mathrm{q}_{m}\|^{2}}\\
	\mathrm{q}_{m}^{\ast} &= p_{m}[\mathbf{u},n] - i r_{1,m}[\mathbf{u},n] - j r_{2,m}[\mathbf{u},n].
\end{align}
Assuming that the orientations are approximately constant in time, the elements in (\ref{eq_coeffnorm}) for $ n = 1,2,\ldots $ can be written as
\begin{equation}
	i(\varphi_{m}^{'}[\mathbf{u},n])\cos(\vartheta_{m}[\mathbf{u}]) + j(\varphi_{m}^{'}[\mathbf{u},n])\sin(\vartheta_{m}[\mathbf{u}])
	\label{eq_phasediff}
\end{equation}
where the term
\begin{equation}
	\varphi_{m}^{'}[\mathbf{u},n] = \varphi_{m}[\mathbf{u},n] - \varphi_{m}[\mathbf{u},n-1]
\end{equation}
is the phase difference. Defining now the unwrapped phase as
\begin{equation}
	\varphi_{m}^{''}[\mathbf{u},n] = \varphi_{m}[\mathbf{u},0] + \sum_{k = 1}^{n}\varphi_{m}^{'}[\mathbf{u},k]  \quad \mathrm{for} \ n = 1,2,\ldots
\end{equation}
the following cumulative sum can be computed
\begin{equation}
	i\varphi_{m}^{''}[\mathbf{u},n]\cos(\vartheta_{m}[\mathbf{u}])  + j\varphi_{m}^{''}[\mathbf{u},n]\sin(\vartheta_{m}[\mathbf{u}]).
	\label{eq_cumsum}
\end{equation}
Filtering the quantity in (\ref{eq_cumsum}) in time, leads to two imaginary quaternionic components 
\begin{equation}
\begin{array}{c}
f_{m}^{(i)}[\mathbf{u},n] = \delta_{m}[\mathbf{u},n]\cos(\vartheta_{m}[\mathbf{u}]) \\
[12pt]
f_{m}^{(j)}[\mathbf{u},n] = \delta_{m}[\mathbf{u},n]\sin(\vartheta_{m}[\mathbf{u}])
\end{array}
\label{eq_filter}
\end{equation}
that define the spatial translation due to a framed motion. In Figure \ref{phase}, the quaternionic phase extraction and unwrapping operations are associated with a single block that is followed by the cumulative sum filtering block.

\paragraph{Signal amplification}
Following the approach presented in~\cite{phasebased}, in order to enhance a motion of interest, the two filtered quaternionic components in~(\ref{eq_filter}) at each pyramid level $m \in \{ 0, \dots, M-1\}$ can be multiplied by the amplification factor $\alpha_{m}$, $m \in \{ 0, \dots, M-1\}$, as shown in Figure~\ref{phase}, obtaining $ \{\alpha_{m}f_{m}^{(i)}[\mathbf{u},n], \alpha_{m}f_{m}^{(j)}[\mathbf{u},n]\}_{m = 0}^{M-1} $.

\paragraph{Signal extraction}
Motion signals can finally be extracted by spatial averaging the amplified and filtered quaternionic components (blocks labelled ``extr.'' in Figure \ref{phase}). Considering a frame of size $U_{1} \times U_{2}$, the following signals are obtained
\begin{equation}
	\begin{array}{c}
	y_{m}^{(i)}[n] = \frac{1}{U_{1}U_{2}} \sum\limits_{u_{1}=1}^{U_{1}-1} \sum\limits_{u_{2}=1}^{U_{2}-1}\alpha_{m}f_{m}^{(i)}[\mathbf{u},n] = \frac{1}{U_{1}U_{2}} \sum\limits_{u_{1}=1}^{U_{1}-1} \sum\limits_{u_{2}=1}^{U_{2}-1} \alpha_{m} \delta_{m}[\mathbf{u},n]\cos(\vartheta_{m}[\mathbf{u}]) \\
	[12pt]
	y_{m}^{(j)}[n] = \frac{1}{U_{1}U_{2}} \sum\limits_{u_{1}=1}^{U_{1}-1} \sum\limits_{u_{2}=1}^{U_{2}-1}\alpha_{m}f_{m}^{(j)}[\mathbf{u},n] = \frac{1}{U_{1}U_{2}} \sum\limits_{u_{1}=1}^{U_{1}-1} \sum\limits_{u_{2}=1}^{U_{2}-1} \alpha_{m} \delta_{m}[\mathbf{u},n]\sin(\vartheta_{m}[\mathbf{u}]).
	\end{array}
	\label{eq_quatcomp}
\end{equation}

\section{Maximum Likelihood Estimation}\label{sec_ml}
Once the motion signals are extracted at each pyramid level, the RR is estimated according to the ML criterion. To this purpose, we first introduce the standard ML principle, that will also be exploited in Subsection \ref{sec_roi} to automatically select ROIs in order to focus on areas where the motion is mainly due to breathing.

The ML principle is indeed a reliable and consolidated method that allows to estimate unknown parameters of interest. Since respiration is characterized by periodic (or quasi-periodic) movements of the chest and abdomen, i.e., expansion and relaxation, the ML criterion can be exploited to detect the presence of a fundamental periodic component, corresponding to the RR, and estimate it \cite{isspit2016}. \\

The RR estimation operation is embedded in the last blocks of Figure~\ref{stvp} and Figure~\ref{phase}. As the motion signals are extracted at each pyramid level for both the presented amplitude- and phase-based approaches, a data aggregation method similar to the one proposed in \cite{multisensors} for multiple sensors can be employed.

For the sake of compactness, the motion signals extracted at each pyramid level can be grouped as follows

	\begin{align}
		\mathbf{l}[n] = &
		\begin{bmatrix}
			\bar{l}_{0}[n]\\
			\bar{l}_{1}[n]\\
			\vdots\\
			\bar{l}_{M-1}[n]
		\end{bmatrix}
		\label{eq_vect_amp} \\
		\mathbf{Y}[n] = &
		\begin{bmatrix}
			y_{0}^{(i)}[n] & y_{0}^{(j)}[n]\\
			y_{1}^{(i)}[n] & y_{1}^{(j)}[n]\\
			\vdots\\
			y_{M-1}^{(i)}[n] & y_{M-1}^{(j)}[n]
		\end{bmatrix}
		\label{eq_vect_ph}
	\end{align}
in the case of amplitude (\ref{eq_vect_amp}) and phase (\ref{eq_vect_ph}) components, respectively. Let us define $\mathbf{X}[n]$ as a generic observation model, that can be written in the form of (\ref{eq_vect_amp}) or (\ref{eq_vect_ph}) according to the considered method. The generic size of $\mathbf{X}[n]$ is $M \times C$, where $M$ is the number of considered pyramid levels and the number of columns $C$ is equal to 1 or 2 in the case of (\ref{eq_vect_amp}) or (\ref{eq_vect_ph}), respectively.

Given the nature of the respiration movements of interest, the observation model  $\mathbf{X}[n]$ is assumed as
\begin{equation}
\mathbf{X}[n] = \mathbf{B} + \mathbf{A} \mathrm{cos} (2 \pi f_{0} T_{s} n + \boldsymbol{\Phi} ) + \mathbf{W}[n]
\label{eq_model}
\end{equation}
where $ \mathbf{B} $ are the continuous components, $ \mathbf{A} $ and $ \boldsymbol{\Phi} $ are the amplitudes and phases, respectively, and $ \mathbf{W}[n] $ are sequences of independent and identically distributed (i.i.d.) zero-mean Gaussian noise samples, all of size $M \times C$. In (\ref{eq_model}), the amplitudes $ \mathbf{A} $, the fundamental frequency $ f_{0} $ and the phases $ \boldsymbol{\Phi} $ are unknown parameters and may be collected as the array of parameters $ \boldsymbol{\Theta} = [\mathbf{A}, f_{0}, \boldsymbol{\Phi} ] $.  Following the standard method presented in \cite[p. 193-195]{Kay} and extending it to the case of multi-dimensional signals, as in \cite{multisensors} and \cite{patwari}, the parameter array $ \boldsymbol{\Theta} $ can be estimated on a window of $N$ frames by minimizing the likelihood function
\begin{equation}
J(\boldsymbol{\Theta}) = \sum_{c=0}^{C-1}\sum_{m=0}^{M-1}\sum_{n=0}^{N-1} \big[ x[m,c,n] -   a[m,c] \, \mathrm{cos} \, (2 \pi f_{0} T_{s} n + \phi[m,c] ) \big]^{2}
\label{eq_likelihood}
\end{equation}
where $  x[m,c,n] $, $ a[m,c] $ and $ \phi[m,c] $ are the generic elements of the matrices $\mathbf{X}[n]$, $ \mathbf{A} $ and $\boldsymbol{\Phi} $, respectively. As shown in \cite[p. 193-195]{Kay}, if the real frequency $f_{0}$ is not close to $ 0 $ or $f_s/2$, an approximate expression of the estimator $\hat{f_{0}}$ of the fundamental frequency can be derived from (\ref{eq_likelihood}) as
\begin{equation}
\hat{f_{0}} = \argmax_{f_{\mathrm{min}} \leq f \leq f_{\mathrm{max}}} \sum_{c=0}^{C-1}\sum_{m=0}^{M-1} \bigg|\sum_{n=0}^{N-1} x[m,c,n] e^{-j2 \pi f T_{s} n }\bigg|^{2}
\label{eq_f0stim}
\end{equation}
where the maximization is performed over the limited frequency interval $[f_{\mathrm{min}}, f_{\mathrm{max}}]$, with $ f_{\mathrm{min}} $ and $ f_{\mathrm{max}} $ being the minimum and the maximum feasible frequencies, respectively, that must be heuristically set.

The amplitudes can similarly be estimated as
\begin{equation}
\hat{a}[m,c] = \frac{2}{N} \sum_{c=0}^{C-1}\sum_{m=0}^{M-1} \left| \sum_{n=0}^{N-1}  x[m,c,n] e^{-j 2 \pi \hat{f}_{0} n T_{s}} \right|
\label{eq_PixelAmplitudesMLV}
\end{equation}
and the presence of a significant periodic component is declared, according to \cite{patwari}, only if the following condition is verified
\begin{equation}
\frac{N}{MC} \sum_{c=0}^{C-1}\sum_{m=0}^{M-1} \hat{a}^{2}[m,c] > \eta
\label{eq_constrampl}
\end{equation}
where $ \eta $ is a properly set threshold.

\subsection{Region of Interest Selection}\label{sec_roi}
\begin{figure}[t!]
	\centering
	\includegraphics[width=\textwidth]{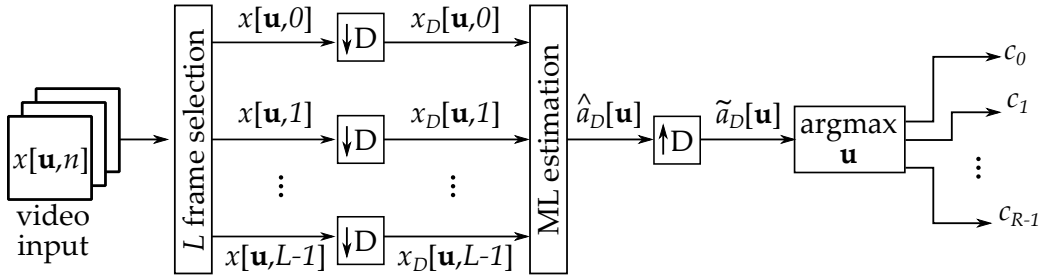}
	\caption{ROI selection algorithm.}
	\label{roi}
\end{figure}

To reduce the computational complexity of the proposed algorithms, a ROI selection algorithm can be exploited to obtain and process video sequences with reduced frame size. In this section, we present an automatic ROI selection algorithm based on the above described ML approach, now applied to the considered video sequence. An illustrative overview of the method is shown in Figure~\ref{roi}.

Given the generic video sequence $ x[\mathbf{u},n] $, the first step for automatically extracting $R$ ROIs consists in selecting $ L $ frames where variations are only due to respiration movements. This processing step is associated with the first block of the diagram in Figure~\ref{roi}. The $L$ frames $ \{x[\mathbf{u},n] \}_{n = 0}^{L-1}$ are first downsampled in space by an integer value $D$ to reduce the computational complexity, obtaining a new block of frames $ \{x_{D}[\mathbf{u},n] \}_{n = 0}^{L-1}$ with a smaller dimension $\lceil U_{1}/D\rceil \times \lceil U_{2}/D \rceil$, where $ \lceil \cdot\rceil $ represents the ceiling operator. This operation is associated with the second bank of blocks of the diagram in Figure~\ref{roi}. The ML approach described in Section~\ref{sec_ml}, associated with the third block of Figure~\ref{roi}, is applied to the downsampled sequences $ \{x_{D}[\mathbf{u},n] \}_{n = 0}^{L-1}$ to estimate the fundamental frequency $ \hat{f_{0}} $ and the amplitudes $ \hat{a}_{D}[\mathbf{u}] $, according to (\ref{eq_f0stim}) and (\ref{eq_PixelAmplitudesMLV}), respectively, where $x[m,c,n]$ and $M \times C$ are now replaced by $x_{D}[\mathbf{u},n] $, i.e., the intensity of the pixel at position $\mathbf{u}$, and $\lceil U_{1}/D\rceil \times \lceil U_{2}/D \rceil$, i.e., the size of the frames.

To compute the centres of the selected $R$ ROIs, the matrix of the amplitudes $ \hat{a}_{D}[\mathbf{u}] $, estimated for the reduced frames, is interpolated at the original frame size $U_{1} \times U_{2}$ to estimate the amplitudes $ \tilde{a}_{D}[\mathbf{u}] $ in the original block of frames.
The centres $ \{c_{r} \}_{r = 0}^{R-1}$ are finally found by selecting the coordinates of the pixels that correspond to the maximum values of $ \tilde{a}_{D}[\mathbf{u}] $. The interpolation and the selection of the ROIs centres are the operations embedded in the fourth and fifth (last) blocks of the diagram in Figure~\ref{roi}, respectively. This procedure allows to extract $R$ ROIs with a fixed size $W \times W$ and may be repeated over time to deal with changes in the position of the framed subject.

\subsection{Large Motion Detection}
To discard ROIs where the motion is affected by large movements unrelated with breathing, a further control procedure may be needed. To this purpose, the intensity of the pixel at position $\mathbf{u}$ at the $n$-th frame of the $r$-th ROI can be defined as $x_{r}[\mathbf{u},n]$ and the pixel-wise difference of consecutive frames can be computed as
\begin{equation}
i[\mathbf{u},n] = x_{r}[\mathbf{u},n] - x_{r}[\mathbf{u},n-1].
\label{eq_firfilter}
\end{equation}
To reduce the computational complexity, the filtered signal in (\ref{eq_firfilter}) could also be binarized according to the following binarization rule
\begin{equation}
i_{r}[\mathbf{u},n]  = \left\lbrace
\begin{aligned}
0& \qquad \text{if } \left|x_{r}[\mathbf{u},n] - x_{r}[\mathbf{u},n-1]\right| < \gamma_{\mathrm{bin}}\\
1& \qquad \text{else}
\end{aligned}
\right.
\quad r = 1,2,\ldots,R
\label{eq:bin}
\end{equation}
where $\gamma_{\mathrm{bin}}$ is  properly chosen binarization threshold.
The average motion signal on the $r$-th region can now be computed as
\begin{equation}
	\bar{i}_{r}[n] = \frac{1}{W^{2}}\sum_{u_{1}=1}^{W-1} \sum_{u_{2}=1}^{W-1} i_{r}[\mathbf{u},n].
	\label{eq_signal2}
\end{equation}
A good decision strategy is such that the $r$-th ROI is discarded if $ \bar{i}_{r}[n] $ in (\ref{eq_signal2}) is above a heuristically set threshold, as expressed by the following decision rule:
\begin{equation}
	\kappa_{r} = \left\lbrace
	\begin{aligned}
	0& \qquad \text{if } \;\bar{i}_{r}[n] > \gamma_{\mathrm{th}}\\
	1& \qquad \text{else}
	\end{aligned}
	\right.
	\quad r = 1,2,\ldots,R
	\label{eq:MotionParameterMLV_k}
\end{equation}
where the binary-valued decision $\kappa_{r}$ defines the presence ($\kappa_{r}$=1) or absence ($\kappa_{r}$=0) of large motion inside the $r$-th ROI and $\gamma_{\mathrm{th}}$ is the selected decision threshold.


Finally, the RR is estimated by maximizing the following likelihood function
\begin{equation}
	J\left( \boldsymbol{\Theta} \right) = \sum_{r=1}^{R}\kappa_{r}J_{r}\left( \boldsymbol{\Theta} \right) 
	\label{eq:RRFusionEstimatorMLV}
\end{equation}
where $ J_{r}\left( \boldsymbol{\Theta} \right) $ is defined according to (\ref{eq_likelihood}) and refers to the $r$-th ROI.

\section{Applications and Results}\label{sec_res}
The performance of the estimation algorithms presented in Section~\ref{sec_motionsignal} and Section~\ref{sec_ml} is now discussed on the basis of experimental results directly obtained by applying the proposed methods on three sets of videos specifically recorded. In particular, the first set includes 2 videos of a newborn \cite{cattani}, whereas the second and third sets include 4 and 10 videos, respectively, of adults sitting still. All videos are recorded indoor by placing a camera laterally or in front of a steady subject normally breathing and not affected by respiratory disorders.

Motion signals are initially extracted and compared with reference data. In the case of the newborn, a pneumogram is used as a gold standard device to acquire the reference respiratory waveform by placing an elastic belt around the chest of the subject. In the case of adults, two wearable sensors, i.e., Shimmer3 by Shimmer Sensing\textsuperscript{TM} and Equivital EQ02 LifeMonitor by Equivital\textsuperscript{TM} are used to record, respectively, the reference accelerometric signal and the respiratory waveform.
A comparison between the proposed amplitude- and phase-based methods is also presented. Results in terms of Root Mean Squared Error (RMSE) between estimated and reference data normalized to the Root Mean Square (RMS) value of the reference data are finally presented.

\begin{figure*}[t!]
	\hspace*{-2.6cm}
	\centering
	\begin{subfigure}{0.6\linewidth}
		\centering
		\includegraphics[width=0.75\linewidth, trim={0 2cm 0 0}, clip]{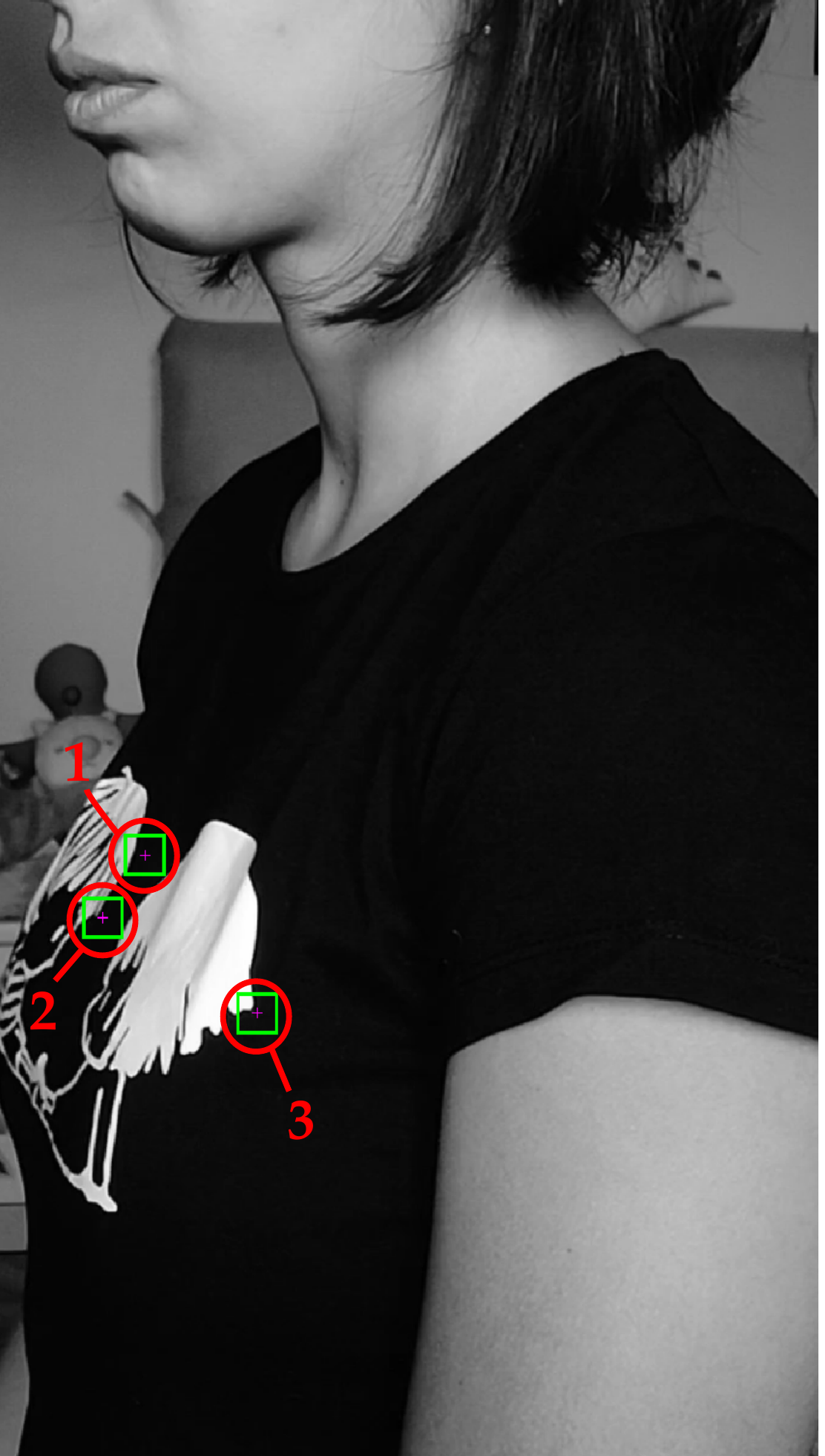}
		\caption{}
	\end{subfigure}%
	\hspace*{-1cm}
	\begin{subfigure}{0.55\linewidth}
		\centering	
		\includegraphics[width=1.5\linewidth]{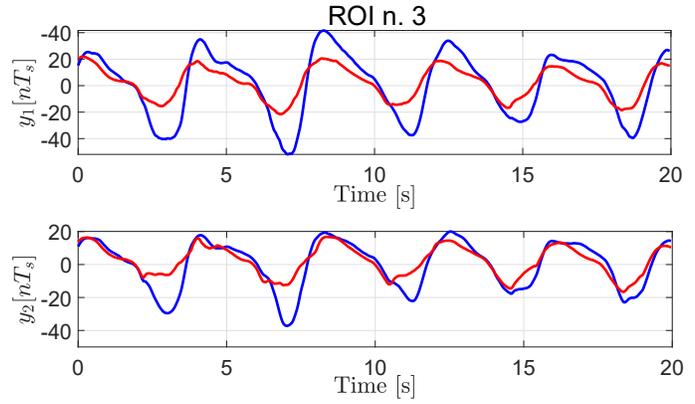} 
		\caption{}
		\vspace*{0.4cm} 
		\includegraphics[width=1.5\linewidth]{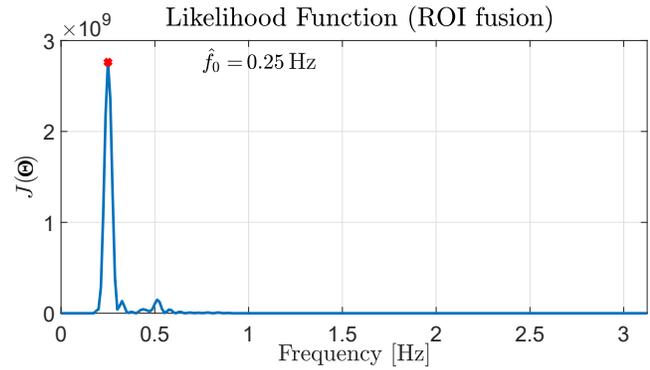}    
		\caption{}
	\end{subfigure}
	\caption{Example of: (a) image of a framed subject where 3 ROIs are highlighted, (b) extracted motion information for $r = 3$ and $m = 1,2$ and (c) likelihood function where the estimated RR is highlighted by the argument of the peak at 0.25 Hz.}
	\label{example1}
\end{figure*}

In Figure~\ref{example1}(a), an illustrative image of a framed subject is shown, highlighting three ROIs as squared regions. The centers of the ROIs are computed according to the procedure detailed in Subsection \ref{sec_roi} for $R = 3$. In Figure \ref{example1}(b), the corresponding motion information extracted by the phase-based motion magnification estimation method is shown.
In particular, the signals $y_{m}^{(\iota)}[n]$, $\iota \in \{i,j\}$, obtained by applying (\ref{eq_quatcomp}), are plotted over a 20~s time window for the third ROI, i.e., $r = 3$, and for two pyramid levels, i.e., $m = 1,2$.
Finally, in Figure~\ref{example1}(c), the corresponding likelihood function obtained by applying~(\ref{eq:RRFusionEstimatorMLV}) is shown as a function of the frequency. The estimated frequency $\hat{f} = 0.25$~Hz is the one corresponding to the maximum peak of the function. 

\begin{figure}[t!]
	\centering
	\includegraphics[width=0.9\textwidth]{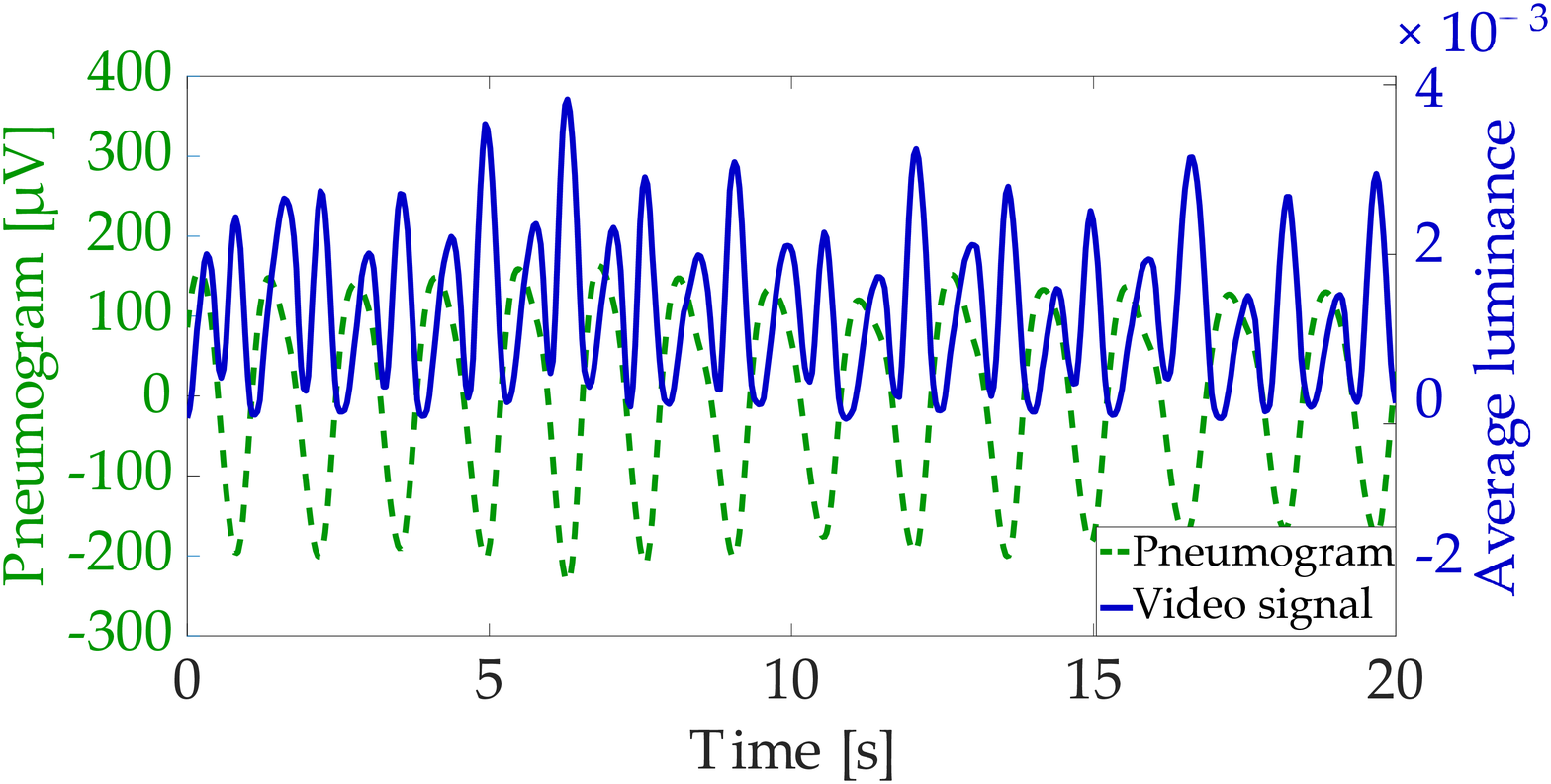}
	\caption{Comparison between the motion signal extracted from a video of a newborn by the amplitude-based motion magnification estimation method and the reference signal, i.e., the pneumogram.}
	\label{refsignal_spazio}
\end{figure}
\begin{figure}[t!]
	\centering
	\begin{subfigure}{0.9\linewidth}
		\centering
		\includegraphics[width=1\textwidth]{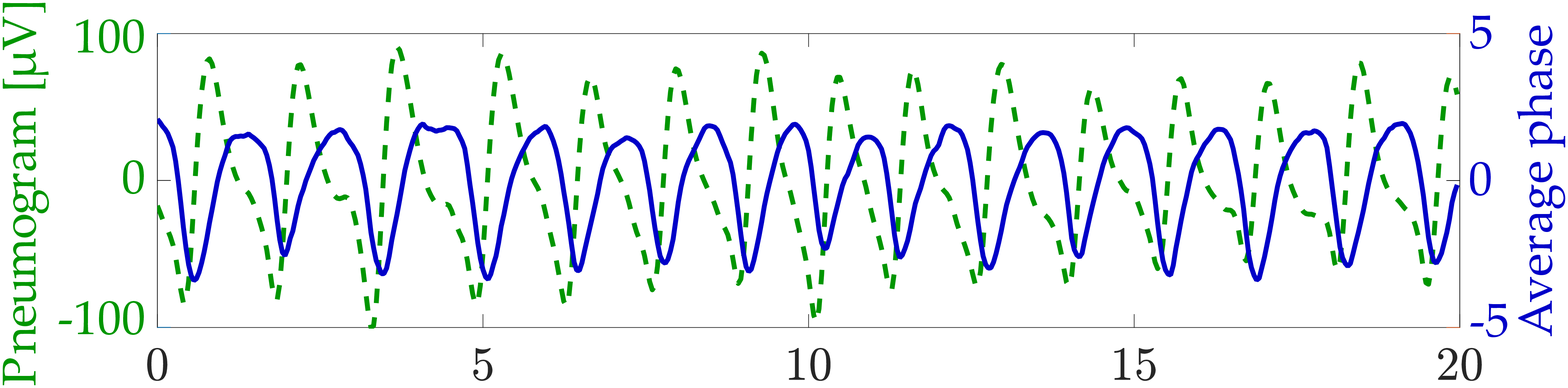}
		\caption{}
	\end{subfigure}
	\begin{subfigure}{0.9\linewidth}
		\hspace*{-0.55cm}
		\centering
		\includegraphics[trim=0cm 0cm 0cm 13cm,clip=true,,width=1.08\textwidth]{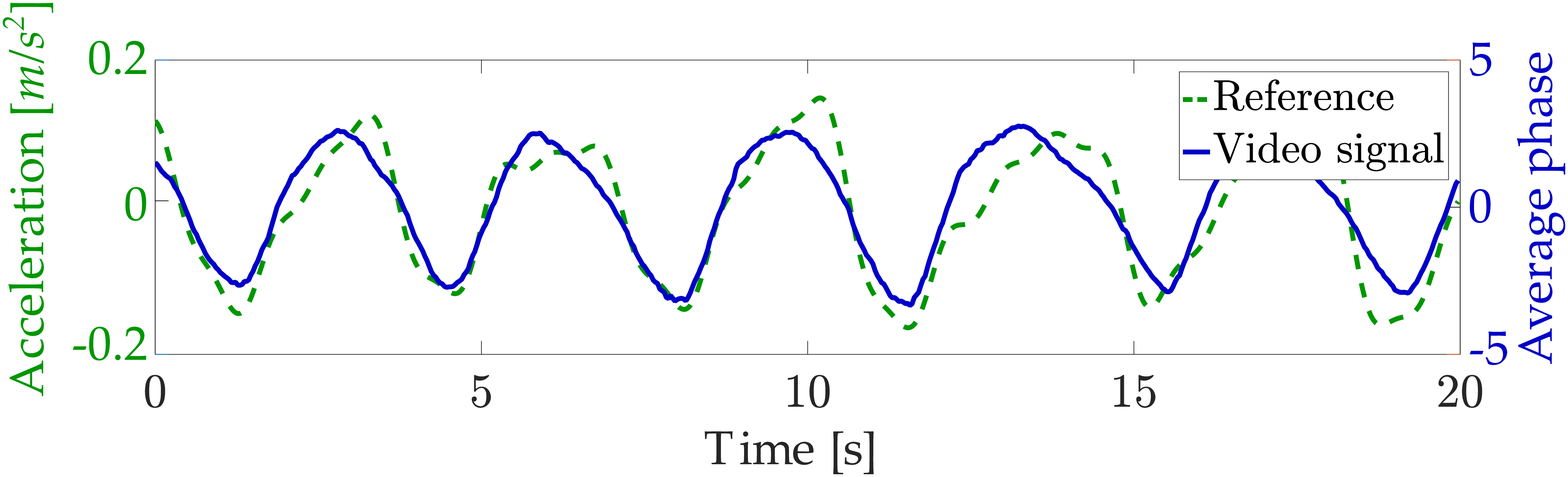}
		\caption{}
	\end{subfigure}
	\caption{Comparison between two motion signals extracted by the phase-based motion magnification estimation method and the reference signals: (a) pneumogram of a newborn, (b) accelerometric signal of an adult.}
	\label{refsignal}
\end{figure}

 
As illustrative examples, motion signals extracted by the amplitude- and phase-based motion magnification estimation methods are shown in Figure~\ref{refsignal_spazio} and Figure~\ref{refsignal}, respectively, along with the corresponding reference signals. In particular, in Figure \ref{refsignal_spazio} the motion signal extracted from a video of a newborn by applying (\ref{eq_signal}) is plotted over a 20 s time window along with the reference signal, i.e., the pneumogram, for the second level ($m = 1$) of the processed pyramid. Considering that one period of the pneumogram corresponds to a complete respiratory cycle, that involves two main movements (inhalation and exhalation), a good correspondence between the two signals can be observed. On the other hand, the average phase variations extracted by two videos, of a newborn and an adult, are plotted over two 20 s time windows and compared with the corresponding pneumogram and accelerometric signals in Figures \ref{refsignal}(a) and \ref{refsignal}(b), respectively. In each case, the two pairs of signals exhibit a comparable periodicity, whereas the differences between the two reference signals, in particular the RR, depend on the employed sensors and on the age of the subject.

\begin{figure}[t!]
	\centering
	\includegraphics[width=0.7\textwidth]{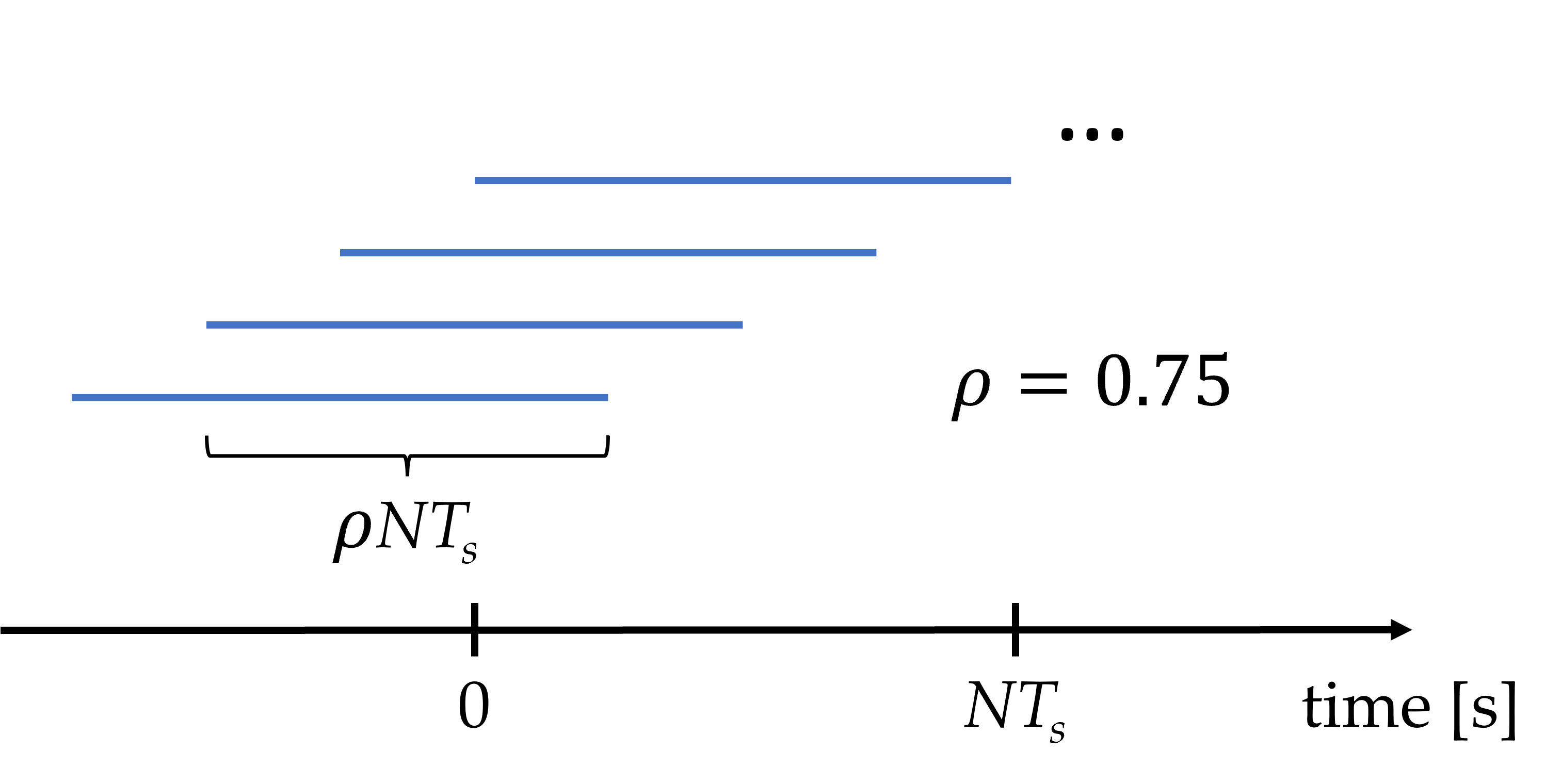}
	\caption{Windows of length $NT_{s}$ s interlaced by a factor $\rho = 0.75$.}
	\label{overlap}
\end{figure}

As further investigation, the ML estimation method presented in Section~\ref{sec_ml} is performed on interlaced windows of $N$ frames, each corresponding to $NT_{s}$~s. In the following results, interlaced windows are considered to track the RR over time with proper resolution and the overlap of consecutive windows is defined by an interlacing factor $\rho \in [ 0, 1 )$. An example of windows of length $NT_{s}$ s interlaced by a factor $\rho = 0.75$ is shown in Figure~\ref{overlap}.

\begin{figure}[t!]
	\vspace*{0.4cm}
	\hspace*{0.3cm}
	\centering
	\includegraphics[width=1\textwidth]{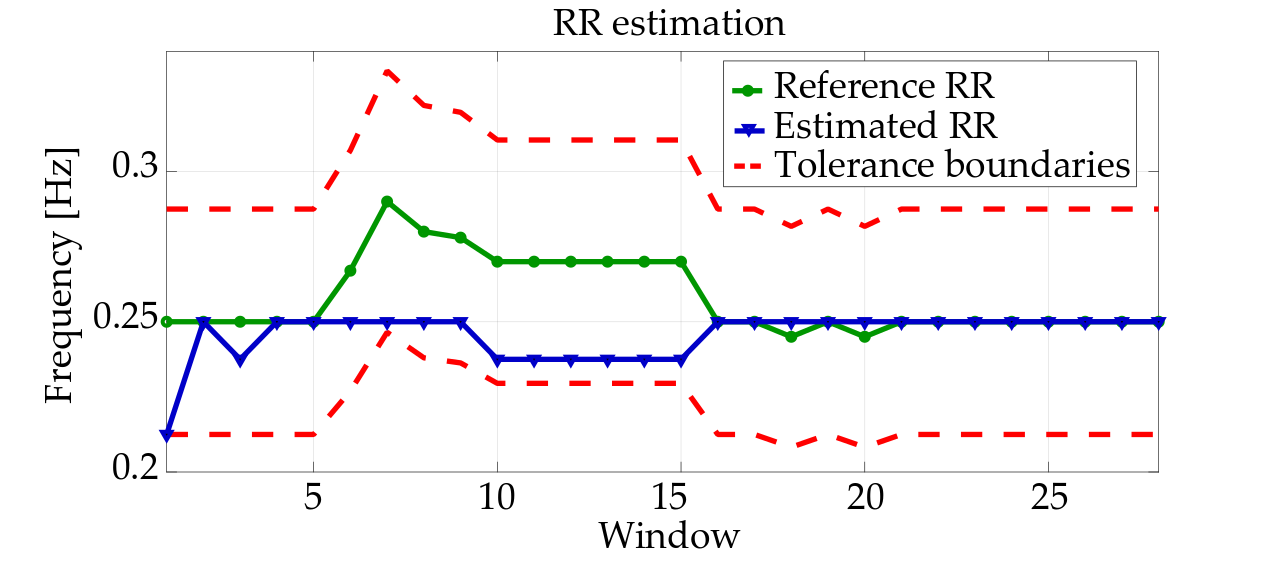}
	\caption{Comparison between estimated and reference RR for the phase-based estimation method.}
	\label{freqcomp}
\end{figure}

In Figure \ref{freqcomp}, the frequencies estimated by the phase-based method on interlaced windows for a video framing an adult sitting still are compared with the reference frequencies estimated by the Equivital EQ02 LifeMonitor. The duration of the considered video is 56 s and the RR estimation is performed on 20 s windows interlaced by $90 \%$ (i.e., $\rho = 0.9$). This corresponds to 28 processed windows. The first 9 windows should not be considered in the analysis because processed data are incomplete due to the chosen window overlap, as shown in Figure~\ref{overlap} for $\rho = 0.75$ which exhibits 3 incomplete initial windows. It can be noticed that the RR is estimated with good approximation in all windows, confirming the robustness of the system. Tolerance boundaries highlighted in Figure \ref{freqcomp} are computed according to the medical practice of considering acceptable a $\pm 15 \% $ variation from to the reference frequency. 

\begin{figure}[t!]
	\centering
	\begin{subfigure}{1\linewidth}
		\hspace*{-0.48cm}
		\centering
		\includegraphics[width=1.05\textwidth]{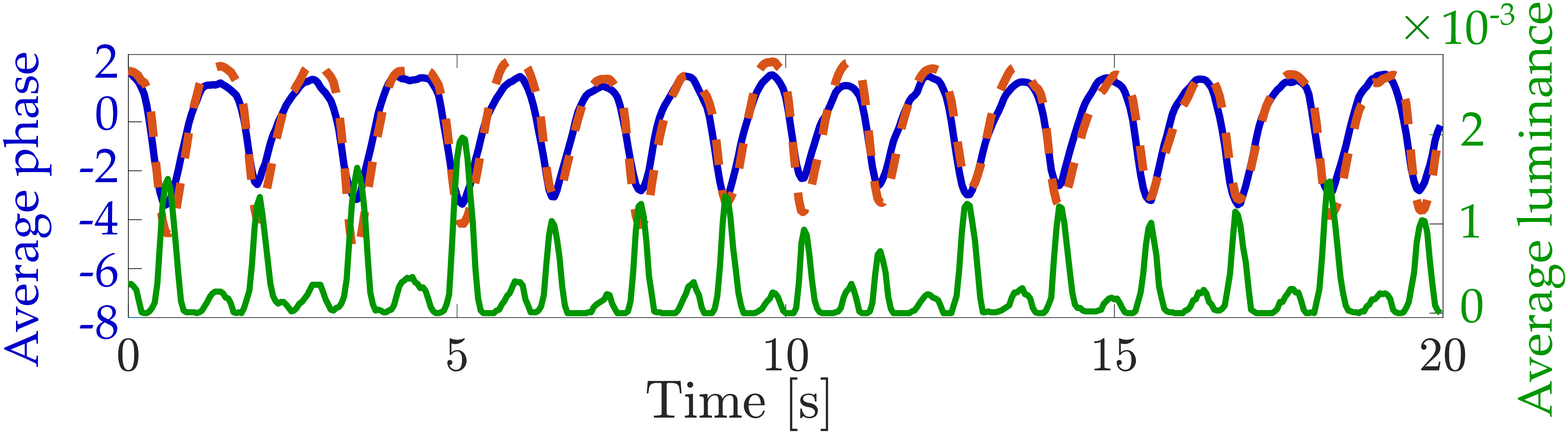}
		\caption{}
	\end{subfigure}
	\begin{subfigure}{1\linewidth}
		\centering
		\includegraphics[width=1\textwidth]{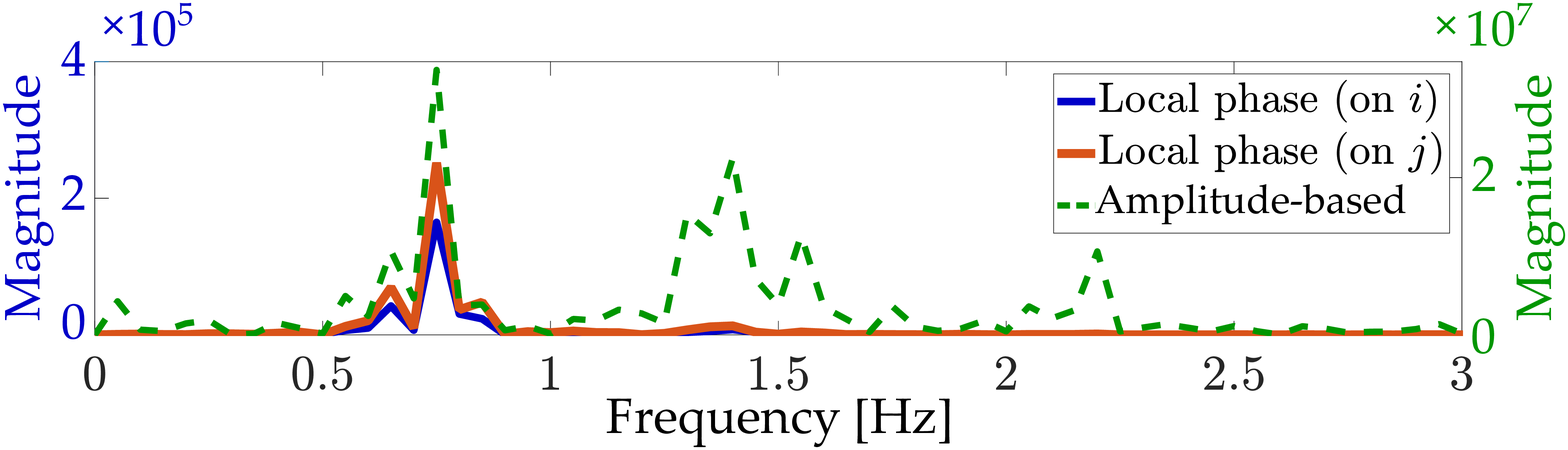}
		\caption{}
	\end{subfigure}
	\caption{Comparison between the presented methods: (a) extracted motion signals and (b) magnitude spectra.}
	\label{comparison}
\end{figure} 
A comparison of the presented amplitude- and phase-based methods is now proposed in Figure~\ref{comparison}. In particular, the signal extracted by the amplitude-based motion magnification estimation method from a video of a newborn is shown in Figure \ref{comparison}(a) along with the corresponding signals $y_{0}^{(\iota)}[n]$, $\iota \in \{i,j\}$, locally extracted from a selected ROI of the considered video by the phase-based method. The duration of the considered video signal is 20 s. The signal extracted by the amplitude-based method is always positive as the quantity obtained by applying (\ref{eq_signal}) defines the average luminance for each processed frame. For this reason, inhalation and exhalation acts, which are characterized by movements in opposite directions, may not be clearly distinguishable, especially under critical conditions, e.g., poor camera positioning or patient type. The phase-based method allows to overcome this limit, as the extracted signals $y_{0}^{(\iota)}[n]$, $\iota \in \{i,j\}$ exhibit negative values, too. The different characteristics of the two types of signals are also visible in Figure \ref{comparison}(b), where their magnitude frequency spectra are plotted. As the phase-based magnification is performed on a selected ROI, the extracted phases are indicated as ``local phases'' in the legend of Figure \ref{comparison}(b). A peak around 0.75 Hz can be observed for all the considered cases, corresponding to the correctly estimated RR of 45~breath/min. Nevertheless, the shape of the signal extracted by the amplitude-based method causes other peaks, related to higher order harmonics, to appear around 1.4~Hz and 2.2~Hz. Under critical conditions, these secondary peaks may be higher than the fundamental one impairing RR estimation: for example, a frequency twice the correct one could be estimated. On the other hand, as the signals extracted by the phase-based method are quasi-sinusoidal, due to the direct application of (\ref{eq_quatcomp}), peaks related to higher order harmonics are negligible. This leads to more reliable RR estimation.

\subsection{Performance Analysis}
To evaluate the performance of the presented methods, videos framing different subjects in different scenarios are analysed. The main characteristics of the considered videos are summarized in Table \ref{tab:RRmethodsCompare}, where the parameter setting for the video processing analysis is also reported. The durations of the videos vary approximately between 1 min 35 s and 5 min. The camera resolution and the sampling frequency vary according to the employed recording device. We recall that the parameters $M$, $W$ and $R$ indicate, respectively, the number of pyramid levels, the fixed size of the ROIs and the number of ROIs. The cut-off frequencies of the employed Butterworth filter, used to extract the frequency band of interest, are denoted as $f_{\mathrm{L}}^{\mathrm{co}}$ and $f_{\mathrm{H}}^{\mathrm{co}}$, $\alpha$ is the amplification factor, $NT_{s}$ is the duration of the processed time window and the interlacing factor $\rho$ denotes the overlap between consecutive estimation windows. The device used as reference is also indicated.

\begin{table*}[t!]
	\renewcommand{\arraystretch}{1.3}
	\begin{center}
		\hspace*{-2.1cm}
		\resizebox{1.3\textwidth}{!}{%
			\begin{tabular}{cccccccccccc}
				\hline
				\bfseries Video& \bfseries No.& \bfseries Camera& $f_{s} $& \multirow{2}{*}{$M$} & $W$& \multirow{2}{*}{$R$} & $[f_{\mathrm{L}}^{\mathrm{co}}, f_{\mathrm{H}}^{\mathrm{co}}]$&  \multirow{2}{*}{$\alpha$} & $NT_{s}$ & \bfseries $\rho$ & \bfseries Reference\\
				\bfseries set & \bfseries samples & \bfseries resolution & \mdseries [Hz] & & \mdseries [pixel] & & \mdseries [Hz] & & \mdseries [s] &  & \bfseries device \\ 
				\hline
				Newborns & 2 & $360\times288$ & 25 & 3 & 21 & 4 & [0.3 1.1] & 25 & 20 & 0.5 & Pneumograph\\
				Adults & 4 & $800\times600$ & 30 & 4 & 41 & 3 & [0.19 0.9] & 20 & 20 & 0.5 & Accelerometer\\
				\multirow{2}{*}{Adults} & \multirow{2}{*}{10} & \multirow{2}{*}{$1920\times1080$} & \multirow{2}{*}{30} & \multirow{2}{*}{3} & \multirow{2}{*}{16} & \multirow{2}{*}{3} & \multirow{2}{*}{[0.19 0.9]} & \multirow{2}{*}{20} & \multirow{2}{*}{20} & \multirow{2}{*}{0.5} & Equivital EQ02 \\
				& & & & & & & & & & & LifeMonitor\\
				\hline
		\end{tabular}}
	\end{center}
	\hspace*{-1.5cm}
	\caption{Characteristics of the considered videos and parameter setting.}
	\label{tab:RRmethodsCompare}
\end{table*}

\begin{figure}[t!]
	\centering
	\includegraphics[width=1\textwidth]{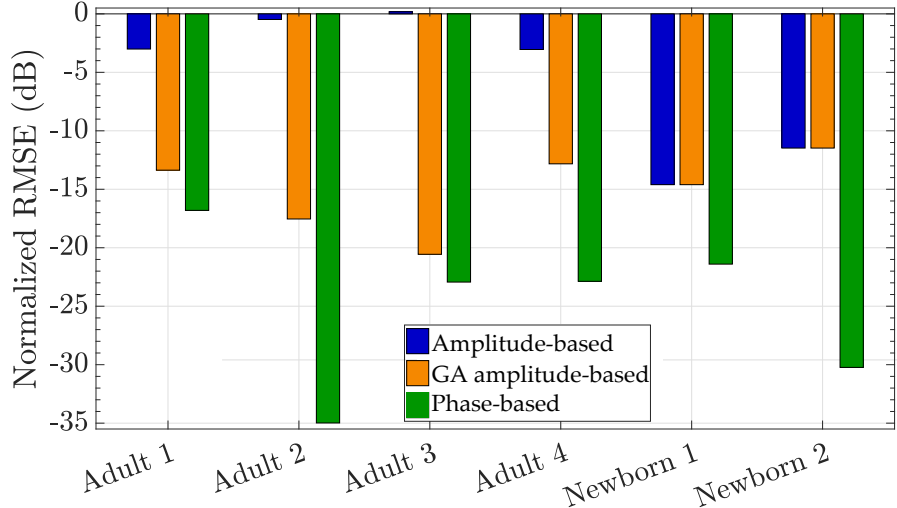}
	\caption{Performance of the assessed methods in terms of normalized RMSE for 6 considered videos (first two sets in Table~\ref{tab:RRmethodsCompare}).}
	\label{resultsRMSE_finaln}
\end{figure}

The accuracy of the presented methods is now analysed in terms of normalized RMSE for 6 tested videos. The results, expressed in dB, are shown in Figure \ref{resultsRMSE_finaln}, where the type of framed subject is reported. Considering $ N_{w} $ temporal windows where the RR estimation is performed, the RMSE for each video is defined as
\begin{equation}
\begin{aligned}
\mathrm{RMSE} = \sqrt{\frac{\sum\limits_{n=1}^{N_{w}}\big|\hat{f}_{0}[n]-{f}_{0}[n]\big|^{2}}{ \sum\limits_{n=1}^{N_{w}} |{f}_{0}[n]|^{2}}}
\end{aligned}
\label{eq_rmse}
\end{equation}
where $\hat{f}_{0}[n]$ and ${f}_{0}[n]$ are the estimated and the reference frequencies for the $n$-th window, respectively. The reference frequencies are obtained by means of an accelerometer, for adults, and a pneumogram, for newborns.
As RR estimation based on the amplitude-based approach is prone to errors caused by higher order harmonics, as previously discussed, an idealized Genie-Aided (GA) version of this method is also considered as a benchmark. The GA method automatically corrects estimated double frequencies. Despite this adjustment, the phase-based method exhibits better performance for all the considered videos. Estimates are indeed more reliable due to the characteristics of the motion signals in (\ref{eq_quatcomp}) of inherently distinguishing motions in the different directions associated with inhalation and expiration.

\begin{figure}[t!]
	\centering
	\includegraphics[width=1\textwidth]{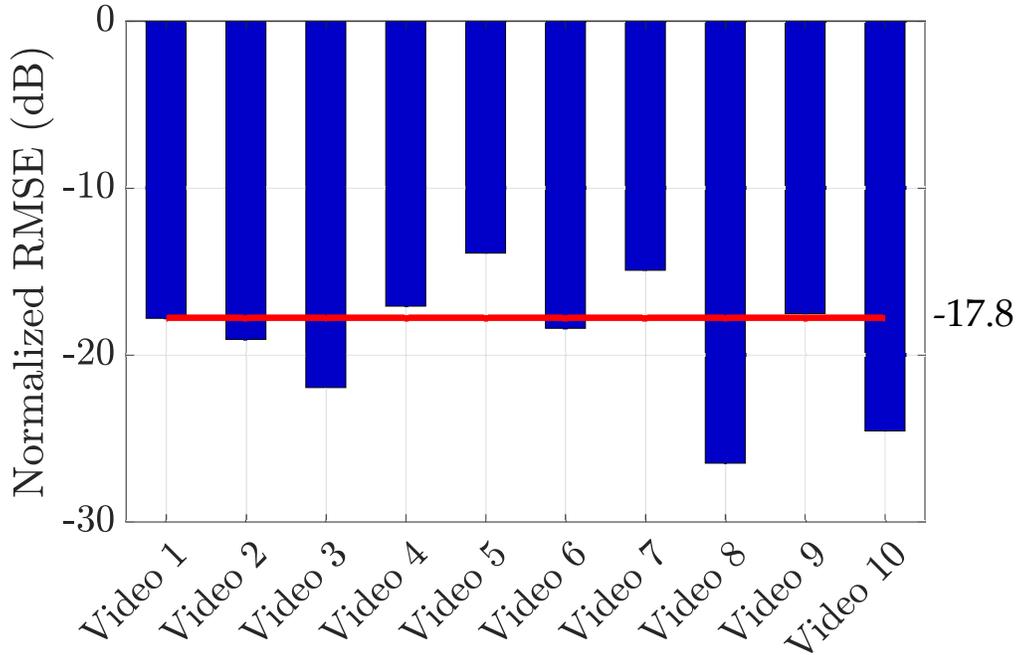}
	\caption{Performance of the phase-based method in terms of normalized RMSE for 10 considered videos (last set in Table~\ref{tab:RRmethodsCompare}).}
	\label{rmse}
\end{figure}

In order to further analyse the performance of the more efficient phase-based method, 10 more videos, all framing adults sitting still, are tested and the normalized RMSE is computed according to (\ref{eq_rmse}). Various subjects, scenarios and camera angles are considered and the Equivital EQ02 LifeMonitor is used as the reference device. The results, expressed in dB, are presented in Figure~\ref{rmse} and show a good agreement with the RMSE values in Figure~\ref{resultsRMSE_finaln}, thus confirming the robustness of the considered method. The average error over all the videos is also highlighted as a straight line at $-17.8$ dB and it can be observed that the RMSE obtained for 6 videos is below or equal to this value. 

\section{Conclusions} \label{conclusion}
In this paper, two contact-less methods to estimate the RR from video sequences are presented. The proposed methods are based on amplitude and phase motion magnification to highlight subtle respiratory movements and combine spatial and temporal processing techniques to extract reliable motion information. Suitable ROIs, where the motion is mainly due to respiration, may be selected to enhance the estimation. Once the motion signals are extracted, the ML principle is applied to estimate, by aggregating data from different ROIs and pyramid levels, the fundamental frequency, corresponding to the RR. The accuracy of the two methods is assessed by comparison with reference data, showing good agreement between the estimated signals and the reference ones. Nevertheless, the characteristics of the motion signal obtained by applying the amplitude-based approach may lead to wrong frequency estimates, doubled with respect to the correct one. These limitations are overcome by the phase-based method that leads to more reliable estimates due to the regular shape of the extracted motion signals which may resemble quasi-sinusoidal ones. The performance of the two methods is compared in terms of normalized RMSE showing the better accuracy of the phase-based approach, that achieves lower errors for all the tested videos. 

\section*{Acknowledgement}
Davide Alinovi received his PhD in 2017 and was a post-doctoral researcher until November 2019, all at the University of Parma. He was a competent, innovative, and passionate researcher who started and carried out the work that led to the preliminary conference papers. Sadly, Davide passed away on September 16, 2020, at the age of 32. His passing left us appalled by the loss of an excellent colleague and dear friend.



	
  \bibliographystyle{elsarticle-num} 
  \bibliography{cas-refs}




\newpage
\end{document}